\documentclass{article}
\usepackage{graphicx}
\usepackage{amsmath}
\usepackage{amsfonts}
\usepackage{amssymb}

\begin{document}

\title{More on A Statistical Analysis of Log-Periodic Precursors to Financial\ Crashes}
\author{James A. Feigenbaum\\Tippie College of Business\\Department of Economics\\University of Iowa\\Iowa City, IA \ 55242\\jfeigenb@blue.weeg.uiowa.edu}
\date{July 2001}
\maketitle
\begin{abstract}
We respond to Sornette and Johansen's criticisms of our findings regarding
log-periodic precursors to financial crashes. \ Included in this paper are
discussions of the Sornette-Johansen theoretical paradigm, traditional methods
of identifying log-periodic precursors, the behavior of the first differences
of a log-periodic price series, and the distribution of drawdowns for a
securities price.
\end{abstract}

\pagebreak 

\qquad In Ref. \cite{Feigenbaum 01}, we reconsidered the evidence regarding
log-periodic precursors to financial crashes. \ We focused much of our
attention on the work of Didier Sornette and Anders Johansen \cite{JS 99}-
\cite{JLS 00}, and they have written a detailed response \cite{SJ 01} to our
paper. \ We now take this opportunity to rebut their criticisms.

\qquad We made four points in \cite{Feigenbaum 01}. \ First, we had some
comments regarding the theory proposed by Johansen \emph{et al} in \cite{JLS
00}. \ Second, we reviewed the traditional curve-fitting approach for
identifying log-periodicity and noted some problems with this methodology.
\ Third and most important, we tested a prediction that comes out of the model
in \cite{JLS 00} that, in addition to the price series, the first differences
of the price series should also behave log-periodically within a log-periodic
precursor. \ In examining the first differences for the S\&P 500 during the
log-periodic spell leading up to the famous October 1987 crash, we found that,
if we restricted attention to the years 1980-1986 (which covers the bulk of
this spell but not the last year prior to the crash), the log-periodic
component of the Sornette-Johansen specification was not statistically
significant at the 5\% level. \ Finally, we weighed in on Johansen and
Sornette's claim \cite{JS 99} that large events like the 87 crash are outliers
in the distribution of drawdowns, a drawdown being the cumulative drop in an
index from a local maximum to the ensuing local minimum. \ We will address
Sornette and Johansen's response to each of these four points separately.\bigskip

\section{The Sornette-Johansen Paradigm\label{paradigm}\bigskip}

\qquad The economic theory proposed by Sornette and Johansen (SJ) to explain
the incidence of log-periodicity in securities prices is described in
\cite{JLS 00} and summarized in Section 2 of our own paper \cite{Feigenbaum
01}. \ Their model assumes that log-periodicity arises from the interaction of
a large group of irrational agents who reside on a network with discrete scale
invariance. \ They also include in their model a single rational agent who is
risk neutral and has rational expectations. \ The introduction of this agent
served two purposes. \ It was intended to defuse the criticism that a
log-periodic bubble could be exploited by rational agents to make unbounded
profits. \ In addition, the no-arbitrage condition derived from this agent's
preferences supplies a mathematical link between the price of the security and
the probability of a crash induced by the irrational agents.

\qquad In their model, they consider a security which earns no dividends, so
any positive price would signify a bubble in the security price $p(t)$. \ The
price can be described in the notation of stochastic calculus as following the
process
\begin{equation}
\frac{dp}{p(t)}=\mu(t)dt+\varepsilon(t)-\lambda dj\text{,}\label{process}%
\end{equation}
where $\mu(t)$ is a time-dependent drift and $\varepsilon(t)$ is a mean-zero
noise term. $\ $The binary variable $j(t)$ keeps track of the impending crash.
\ It will equal zero before a crash and one afterwards, so $dj$ will vanish
except at the instant of the crash, at which time the price of the security
will fall by the factor $\lambda$. \ (A more general model would have
$\lambda$ be the mean of a distribution of possible crash sizes.) \ The
no-arbitrage condition is then a martingale condition on the price:
\begin{equation}
E[dp]=0\text{.}\label{martingale}%
\end{equation}

\qquad One observation made in \cite{Feigenbaum 01}, which was also made by
Ilinski \cite{Ilinski 99}, is that such a no-arbitrage condition precludes the
possibility of making positive profits with certainty. \ That is why it is
called a no-arbitrage condition. \ At this point, it is important to note that
what an economist means when using the term profit differs from what an
accountant or the typical layperson would mean. \ Economic profit is the
surplus of revenue over opportunity cost, opportunity cost being the stream of
income that is lost because one did not allocate the resources in question to
some other purpose. \ In this context, the opportunity cost will be the income
lost because one did not invest one's resources in the best possible
alternative manner. \ When we say that the no-arbitrage condition precludes
the possibility of making a certain, positive profit, we do not mean that
investors can earn no revenue from trading in this security. \ Rather, we mean
that an agent cannot improve on a strategy which earns zero expected return.

\qquad In \cite{SJ 01}, SJ argue incorrectly that, in fact, positive profits
could be earned by a savvy investor in their model. \ They make a distinction
between average returns and conditional returns. \ The no-arbitrage condition
(\ref{martingale}) requires that the average unconditional return be zero.
\ However, the return conditional on no crash having occurred, represented by
$\mu(t)$, can and generally will be nonzero.

\qquad This is true, but they are missing the point. \ Investors are not
interested in conditional returns. \ They cannot depend on a crash not
happening. \ If they could simply ignore the possibility of a crash, there
would be no bubble. \ So yes, it is possible that, after the fact, an investor
might find he has earned some money by trading in this security. \ However,
before the fact, he will have just as much expectation that he might lose that
same amount. \ There is no way to construct a trading strategy that will
exploit this bubble to earn an expected return greater than zero. \ That is
the upshot of the no-arbitrage condition. \ If one is going to presume that
the Sornette-Johansen paradigm is valid--keeping in mind we do not advocate
that presumption--then it would be quite irrational to attempt to devise a
money-making strategy which uses log-periodicity to forecast crashes. \ One
might as well spend one's time seeking to invent a perpetual motion machine.

\qquad As SJ note \cite{SJ 01}, there is some empirical evidence that the
efficient market hypothesis is false and that it might be possible to earn
positive profits relative to a standard diversification strategy. \ However,
this evidence does not support their model. \ On the contrary, it goes against
the assumptions of their model.

\qquad Another point which Ilinski \cite{Ilinski 99} made that we expanded
upon in \cite{Feigenbaum 01} was the criticism that SJ's model is based on the
unrealistic assumption that only a single rational agent exists, who also
happens to be risk neutral. \ In \cite{SJ 01}, SJ make a fair stab at
loosening the risk neutrality assumption. \ They draw upon a stochastic
discount factor (SDF) model in \cite{Cochrane 01} to broaden their results and
conclude that ``the most general form of risk aversion does not invalidate our
theory.'' \ As anyone familiar with the literature on expected utility theory
should know, Cochrane's model, while certainly useful, is by no means an
exhaustive model of risk averse preferences. \ However, this specification is
presumably general enough to serve the purpose of showing their model can
accomodate the possibility that their single rational agent is risk averse.

\qquad Unfortunately for SJ, the multiplicity issue will be much more
difficult to tame. \ They note in passing \cite{SJ 01} that ``the SDF is not
different from one agent to the next . . . because it describes the aggregate
perception by the rational agents of the level of risks.'' \ Essentially, SJ
are invoking what is known as a representative-agent model of the
macroeconomy. \ Under certain conditions, such as complete markets, it can be
shown that \emph{in equilibrium} an economy of many agents can be replaced by
a single imaginary agent, and, indeed, the stochastic discount factor of all
the individual agents would equal the SDF for this imaginary agent
\cite{Constantinides 82}.

\qquad The problem with this argument is that it is circular reasoning, for it
requires than an equilibrium consistent with the bubble exists. \ They are
assuming what they are supposed to be proving. \ One purpose of introducing
the rational agent was to forestall the criticism that a rational agent might
exploit a log-periodic bubble to earn unbounded profits, a situation that
could not occur in equilibrium and has never been observed empirically.
\ Quoting from Johansen \emph{et al} in \cite{JLS 00}, ``Lest this sound like
voodoo science, let us reassure the reader that the ability to predict the
critical date is perfectly consistent with the behavior of the rational agents
in our model: they all know this date, the crash may happen anyway, and they
are unable to make any abnormal risk-adjusted profits by using this
information.'' \ Thus far, we are not reassured.

\qquad It is straightforward to construct a price process consistent with
equilibrium for one rational agent, but it is considerably more difficult to
accomplish this for two or more rational agents. \ Allowing an exotic price
bubble compounds the situation further. \ It is more than likely that very
stringent assumptions would have to be put on agents' preferences to construct
a market equilibrium with a log-periodic price bubble.

\qquad Of course, one could dispense with rationality and no arbitrage, and
then it is trivial to get log-periodicity in price processes as Sornette and
Ide demonstrated in \cite{SI 01}. \ The trick is to combine all these
elements, and so far no one has accomplished this except under highly
unrealistic assumptions.\bigskip

\section{Fitting the Price Curve\label{fitting}\bigskip}

\qquad The majority of \cite{SJ 01} is devoted to attacking the third section
of \cite{Feigenbaum 01} in which we review past work on identifying
log-periodicity. \ In their words, ``We are obviously not going to divulge our
technique and methodology for crash prediction but instead offer a few common
sense guidelines to avoid the rather obvious traps in which Feigenbaum has
fallen.'' \ One could just as well argue that they are the ones falling into
the traps. \ We see only a marginal role for common sense in determining their
guidelines. \ Their rules of thumb appear to have been chosen to produce the
results they want to show. \ They give no econometric justification for
rejecting our own inferences.

\qquad The specification that we used in \cite{Feigenbaum 01} came from
\cite{SJ 97}. \ The log of the price is assumed to obey
\begin{equation}
\ln(p(t))=A+Bf_{1}(t)+Cf_{2}(t),\label{logspec}%
\end{equation}
where
\[
f_{1}(t)=\frac{(t_{c}-t)^{\beta}}{\sqrt{1+\left(  \frac{t_{c}-t}{\Delta_{t}%
}\right)  ^{2\beta}}}%
\]
and
\[
f_{2}(t)=f_{1}(t)\cos\left[  \omega\ln(t_{c}-t)+\frac{\Delta_{\omega}}{2\beta
}\ln\left(  1+\left(  \frac{t_{c}-t}{\Delta_{t}}\right)  ^{2\beta}\right)
+\phi\right]  .
\]
This is a specification with three linear parameters $A$, $B$, and $C$; and
six nonlinear parameters $\omega$, $\beta$, $\Delta_{t}$, $\Delta_{\omega}$,
$t_{c}$, and $\phi$. \ We should note that the critical time $t_{c}$ does not
correspond to the time of the crash but, rather, to a time when the
probability of a crash is maximized.

\qquad Econometrically, this is an extremely unfriendly model because there
are so many nonlinearities. \ Furthermore, in the Sornette-Johansen paradigm
this specification is supposed to correspond to an integral of the function
$\mu(t)$ in Eq. (\ref{process}), which would imply that $\ln p(t)$ is
autocorrelated. \ As we demonstrated with a pedagogical example in Section 4
of \cite{Feigenbaum 01}, regressing an autocorrelated time series on functions
of time alone can lead to spurious results, and this kind of regression is
precisely what is being done when one fits a stock index to the specification
(\ref{logspec}).

\qquad The many curve fits of log-periodic spells existing in the literature
are producing descriptive statistics regarding the realization of price
trajectories. \ However, these descriptive statistics may offer no information
about the structural parameters of the underlying stochastic process. \ SJ's
primary objection \cite{SJ 01} to our results in \cite{Feigenbaum 01} is that
we often obtain values for the nonlinear parameters which are either too large
or too small relative to some unspecified standard. \ In the absence of any
real log-periodic structure in the data, there is no basis for expecting the
nonlinear parameters to satisfy their criteria. \ If we blindly apply a
complicated regression specification to a data set, we will necessarily obtain
parameter estimates, but these estimates will be meaningless if the
specification is wrong.

\qquad SJ put great faith in the values of $\beta$ and $\omega$ that they find
in their many examples of log-periodic spells, and they are quite impressed
that they always manage to get similar values. \ However, we do not replicate
this result with our own procedure. \ For example, our best fit for the 80-87
data set produces very different values of $\beta$ and $\omega$ from what
Johansen \emph{et al} report in \cite{JLS 00}. \ One possible explanation for
this discrepancy, suggested by their criticism mentioned above, is that they
are constraining their fitting procedure to consider only values of the
nonlinear parameters which fall into a chosen set. \ If this is what they are
doing, it would generally be considered inappropriate and would cloud the
issue of whether they are actually estimating the population parameters they
believe they are estimating.

\qquad Regarding the Lomb periodograms of \cite{JLS 99}, we erroneously stated
in \cite{Feigenbaum 01} that this procedure offered an independent means of
estimating the frequency of a log-periodic fit. \ In fact, the method
described in \cite{JLS 99} is not independent from their curve-fitting
procedure since they use the fit to determine $t_{c}$ and to detrend the data.
\ It would be very surprising indeed\ if they found a low power at the
frequency $\omega$ after they have already shown they can fit the data to an
oscillation with frequency $\omega$. \ Consequently, the consistency of their
Lomb periodograms with their curve fitting does not help to corroborate SJ's claims.

\qquad In \cite{SJ 01},\ SJ also express some outrage regarding our dismissal
of previous statistical analyses carried out by Johansen \emph{et al}
\cite{JLS 99}-\cite{JLS 00} and by Feigenbaum and Freund \cite{FF 96}. \ We
stand by our comments from the first paper. \ With respect to Monte Carlo
simulations, we can only use them to establish statistical properties of a
stochastic process if the data-generating process (DGP) used in these
simulations shares those properties. \ The fact that simulations produce
behavior different from what we observe in real markets only proves that the
DGP driving these simulations is not the true DGP.

\qquad With respect to studies involving fits of out-of-sample data, many of
SJ's supposed fits are not impressive. \ Just leafing through the figures in
\cite{SJ 01}, we find examples where the alleged log-periodic oscillations are
dwarfed by the volatility in the data and it is very hard to believe we are
looking at the best fit to satisfy the specification (\ref{logspec}), which
may not even be a correct specification. \ Financial economists can tell many
stories of the folly in putting too much faith in patterns discovered by
poring through out-of-sample data. \ If log-periodicity is truly a harbinger
of crashes, this is not going to be established with data mining. \ Its power
as a forecasting tool can only be measured by its ability to forecast. \ We
must wait to see what the future brings to judge this issue.

\qquad That said, SJ do present one interesting new piece of evidence in
\cite{SJ 01}. \ They describe the result of a systematic scan of data for the
Hang-Seng index using a moving window of 1.5 years from 1980 to 1999. \ During
this period, they found 9 examples of log-periodic spells that satisfy their
criterion. \ All but one of these was followed by a drawdown of 5\%\ or more,
a result so good that it should make one suspicious. \ This was also a period
in which they claim at least eight (independent) drawdowns of 10\% or more
occurred. \ Clearly, the Hang-Seng is a highly volatile index. \ The average
time between these large drawdowns is about 2.5 years, about the same order of
magnitude of their window size.

\qquad Now, of course, SJ would argue that this is great evidence of the power
of log-periodicity to predict crashes. \ But they give no information about
how well the estimation of $t_{c}$ using their procedure does at estimating
the actual time of the ensuing crash, and this suggests another mechanism
might be at work here. \ It is possible that their selection criteria will,
with great probability, rule out time periods in which a large drawdown
occurs. \ If that is the case, their procedure may simply be flagging periods
in which crashes do not occur. \ And since crashes are very frequent in this
20-year interval, the time between a flagged period and the next crash is
likely to be short, so it may look like their procedure is forecasting crashes
when it is not really doing that at all.\bigskip

\section{Behavior of First Differences\label{FD}\bigskip}

\qquad The most important section of \cite{Feigenbaum 01}, where we made our
main original contribution, received the least attention from SJ in \cite{SJ
01}. \ As we already noted in the previous section, if we assume $\ln p$ is
following a process of the form (\ref{process}) then least squares estimators
of the parameters in Eq. (\ref{logspec})\ can give spurious results if one
regresses $\ln p$ directly on functions of time. \ Instead, the appropriate
procedure is to look at the first differences of $\ln p$. \ We did this in
Section 4 of \cite{Feigenbaum 01} for the log-periodic spell in the S\&P\ 500
leading up to the October 1987 crash.

\qquad As we discussed in the previous paper, the least-squares objective
function for the specification (\ref{logspec}) does not have a clearly defined
global minimum with respect to the nonlinear parameters, so it is not clear
how one should properly estimate the mean and, more importantly, the variance
of least-squares estimators for these nonlinear parameters. \ We, therefore,
focused on the linear parameters $A$, $B$, and $C$. \ It is well accepted in
the finance literature that the noise-term $\varepsilon(t)$ in (\ref{process})
is dependent on previous values of $\varepsilon$, which would invalidate
standard error formulas for the estimation of $A$, $B$, and $C$. \ (SJ
\cite{SJ 01} insinuate that, because the distribution of $\varepsilon$ is
fat-tailed as opposed to Gaussian, least squares methods are inappropriate for
estimation in this context. \ However, consistent estimation with least
squares methods does not require Gaussian noise distributions, which are by
far the exception in econometrics.) \ To overcome the problem of serial
dependence, we estimated critical values for the $T$ statistics of the linear
parameters using a Monte Carlo simulation. \ This procedure requires that we
specify a DGP to model the behavior of the S\&P\ 500. \ We used a random walk
for this purpose. \ While we do not believe the S\&P 500 actually follows a
random walk, we can set the parameters of the random walk to match some of the
properties of the S\&P 500, in this case the first two moments.

\qquad In \cite{SJ 01}, SJ make note of an apparent inconsistency here. \ In
\cite{Feigenbaum 01}, we dismissed their Monte Carlo results because they only
considered one possible DGP for securities prices, a GARCH model. \ Yet here,
we expect them to accept our own Monte Carlo results which also consider only
one DGP.

\qquad There is a subtle point of reasoning that SJ evidently missed here.
\ In their case, they are trying to persuade readers to accept an alternative
hypothesis. \ In contrast, we were testing and rejecting such a hypothesis.
\ As they admit, ``No truth is ever demonstrated in science; the only thing
that can be done is to construct models and reject them at a given level of
statistical significance.'' \ In classical statistics, one either rejects a
hypothesis or one fails to reject it. \ In their case, they rejected the null
hypothesis that a\ GARCH\ model could produce the behavior they observed in
real markets. \ If their experiment had turned out the other way, we would
probably not be writing this paper because people would have concluded there
was nothing extraordinary about log-periodic fits. \ Instead, the experiment
went as they hoped, and we can conclude that the actual stochastic process is
not a GARCH process. \ That is all we can conclude. \ We cannot draw any
conclusions about the vali\-dity of log-periodicity because the space of
possible DGPs is huge, and all they ruled out is an infinitesimal fraction of
it. \ Note also that the fact that the GARCH model is ``one of the most
fundamental benchmarks of the industry'' is irrelevant. \ The GARCH model is a
very good model of how stock prices behave under normal circumstances, but
SJ's own investigations demonstrate how poorly it captures the behavior of
stock prices under the extreme circumstances surrounding a large drawdown.

\qquad In contrast to their rejection of the GARCH, we failed to reject the
null hypothesis that a random walk could produce the log-periodic behavior we
were looking at. \ We can, therefore, conclude that the random walk-like
properties of the true DGP\ are enough to produce this behavior. \ It is
reasonable to expect that other DGPs with similar properties can also
replicate this behavior. \ Whereas SJ have merely eliminated from
consideration one out of an infinite number of possible DGPs, we have
established that the observed log-periodic behavior in first differences is
not extraordinary. \ Thus, the negative result of our paper speaks much louder
than the positive results that they and others have found.

\qquad We tested the specification
\begin{equation}
\Delta\ln p(t_{i})=0=A+B\Delta f_{1}(t_{i})+C\Delta f_{2}(t_{i})+\sum
_{s=2}^{4}D_{s}\delta_{si}+\varepsilon_{i},
\end{equation}
where we had closing prices sampled at $N$ dates $t_{1},\ldots,t_{N}$ and
$\varepsilon_{i}$ is a mean-zero noise term. \ The covariate $\delta_{si}$ is
a delta function equal to 1 if $\Delta t_{i}=s$ and 0 otherwise. \ Note that
in the Sornette-Johansen paradigm the constant $A$ and the time dummy
variables $D_{2}$, $D_{3}$, and $D_{4}$ will vanish. \ One criticism of SJ's
work is that they have done no specification tests. \ They assume their
hypothetical specification is the correct one and do not consider the
possibility that an alternative might do better, which would constitute a
rejection of their hypothesis. \ The inclusion of these extra terms here
allows for the possibility that the upward trend seen in most stock prices
leading up to a drawdown results from a constant drift rather than the more
complicated $f_{1}$ term of their specification.

\qquad For the period from January 1980 to September 1987, we found that $B$,
the coefficient of $\Delta f_{1}$, was not statistically significant. \ Thirty
percent of our Monte Carlo simulations produced a $T$ statistic $B$ equal or
larger in magnitude than what we obtained for our best fit. \ In \cite{SJ 01},
SJ dismiss this finding, attributing the insignificance of $f_{1}$ to ``the
intrinsic difficulty in quantifying a trend and an acceleration in very noisy
data.''\ \ However, one could invoke similar difficulties to dismiss the
entire field of log-periodic precursors.

\qquad For $C$, the coefficient of the log-periodic term, we found that it was
statistically significant at the 5\%\ level if we considered the whole data
set. \ However, if we restrict our attention to the period from 1980-1986,
twenty-five percent of our Monte Carlo simulations produced a $C$ equal or
larger in magnitude to the result of the best fit. \ Thus, we concluded that
the log-periodic component was not statistically significant for the bulk of
the log-periodic spell preceding the 87 crash. \ In \cite{SJ 01}, SJ dismiss
this result also:\ \ ``It is as if a worker on critical phenomena was trying
to get a reliable estimation of $t_{c}$ and $\beta$ by thrashing the last 15\%
of the data, which are of course the most relevant.'' \ This argument
presupposes that we are, in fact, looking at a critical phenomenon, which in
our opinion has yet to be established. \ In any case, that is beside the
point. \ If it was our purpose to measure nonlinear parameters like $t_{c}$
and $\beta$, then clearly it would be a mistake to throw out the data where
the critical behavior is most pronounced. \ But that was not our purpose.
\ Our purpose was to determine whether these functions had a statistically
significant presence throughout the data, and the answer is they do not.
\ Granted, this does not prove that log-periodicity is absent from the first
difference series during these six years. \ It only proves we cannot
convincingly detect log-periodicity there. \ Nevertheless, given this null
result, a theory which can explain the log-periodic behavior in $\ln p$
without requiring log-periodic behavior in $\Delta\ln p$ will be preferred
over the Sornette-Johansen paradigm by Occam's Razor.

\qquad Before moving on, we must point out that SJ are being hypocritical when
they chastise us for throwing out the last year of data. \ As we noted in
\cite{Feigenbaum 01}, it is common practice in this literature to throw out
from these curve-fitting procedures the end data points closest to the crash,
sometimes eliminating a period as large as a few months. \ SJ even confess to
this practice in \cite{SJ 01}. \ These end data points are even closer to the
alleged critical point than most of the data we excluded. \ If these are the
crucial data points, then we have all committed a great sin.\bigskip

\section{The Frequency Distribution of Drawdowns\label{drawdown}\bigskip}

\qquad The last subject considered in \cite{Feigenbaum 01} was SJ's claim,
first put forward in \cite{JS 99} and later elaborated upon in \cite{JS 00},
that large events like stock market crashes are outliers in the frequency
distribution of drawdowns. \ They maintained that all but the largest
drawdowns fit well to an exponential or a stretched exponential distribution,
and they, therefore, posited that a different mechanism must be producing
these outlier events. \ We responded to this claim by doing a specification
test on this drawdown distribution, and we found both the exponential and the
stretched exponential specifications could be rejected.

\qquad In \cite{SJ 01}, SJ answer this challenge with three new pieces of
evidence. \ First, they describe the results of a bootstrap test. \ They
reshuffled the sample of daily returns for the NASDAQ composite index from
1971 to 2000 and generated 10000 days of simulated data. \ The resulting
drawdown distribution produced a much smaller frequency of large events than
is observed in the real data. \ However, this is not surprising. \ They are
bootstraping the unconditional distribution of daily returns, but it is well
established that daily returns are dependent on previous returns, and, in
fact, the stock market may even be a long memory process. \ All of these
dependencies are ignored by this bootstrap test. \ We can conclude from this
test that these dependencies, indeed, play an important role in producing
large drawdowns, but this does not establish that large drawdowns are outliers.

\qquad Next, they describe the drawdown distribution resulting from Monte
Carlo simulations of the stock market as a GARCH process. \ This simulated
distribution also differs from the empirical distribution, but again this only
substantiates their result that the true DGP\ underlying securities prices is
not a GARCH\ process.

\qquad Finally, they corroborate our result that the stretched exponential
does not fit well to moderate drawdowns larger than 5\%. \ But instead of
abandoning the claim that large drawdowns are outliers, they expand the class
of drawdowns that they consider to be outliers. \ To some extent, we must
recognize that this is an argument over semantics. \ What do we mean by an outlier?

\qquad Moving beyond semantics, we do not agree with SJ's claim that ``the
exponential distribution is the natural null hypothesis for uncorrelated
returns.''\ \ In \cite{JLS 00}, Johansen \emph{et al} derive the result that
the drawdown distribution should be exponential under the same assumption of
their bootstrap test that changes in a stock price from one instant to the
next are independent. \ This assumption is certainly false, though it may be a
reasonable approximation for small drawdowns. \ As the length of a chain of
similarly signed innovations increases, however, the serial dependence in this
chain will likely become more important. \ Thus, while it may be a natural
null hypothesis that the distribution should be exponential in the limit of
small drawdown sizes, it is not fair to extrapolate this hypothesis to large
drawdown sizes. \ Moreover, this weaker null hypothesis that the distribution
be exponential only in the small drawdown limit is supported by the
preponderance of the evidence.

\qquad We also quibble with their claim that one could not ``prove'' that
large drawdowns are outliers. \ As we discussed in \cite{Feigenbaum 01}, if
one could reliably distinguish drawdowns with a log-periodic precursor from
drawdowns lacking such a precursor, then one could determine the distribution
of these two populations of drawdowns. \ If the distributions differ in a
statistically significant manner, that would go a long way to establishing the
claim that a different mechanism is responsible for the two kinds of drawdowns.\bigskip

\section{Concluding Remarks\label{conclusion}\bigskip}

\qquad Speaking as a researcher who has been involved from the beginning in
the study of log-periodicity in the stock market, we believe that the initial
evidence of crashes as critical phenomena was certainly intriguing. \ However,
it is important not to get carried away by a new hypothesis and lose one's
scientific objectivity. \ SJ lament \cite{SJ 01} that ``it is an all too
common behavior to dismiss lightly a serious hypothesis by not taking the
trouble to learn the relevant skills necessary to test it rigorously.''
\ However, they themselves have thrown roadblocks in the way of testing it
rigorously. \ As long as they keep the details of their methodology secret,
their results cannot be independently tested or reproduced. \ They have also
disregarded many econometric procedures that would generally be considered the
most appropriate tools for addressing the problems they encounter. \ Instead,
they cobble together methods with little or no theoretical basis and then they
question why nobody follows their trail. \ Furthermore, when one finds that
standard methods produce results contrary to their hypothesis, one has to
consider if maybe they have rejected these standard methods precisely because
the results obtained do not support their hypothesis.

\qquad SJ also draw an analogy between log-periodicity and global warming.
\ We think a better analogy can be made to the \emph{canalis} of Mars. \ For
over a hundred years, astronomers like Giovanni Schiaparelli and Percival
Lowell were almost fanatical in their belief that they could see evidence of a
great Martian civilization through their telescopes. \ Eventually, telescopic
resolutions improved to the point where astronomers could clearly see there
was no system of artificial canals crisscrossing the surface, yet it took a
long time for the idea to die.\bigskip

\noindent{\Large \textbf{Acknowledgements\bigskip}}

\qquad We would like to thank Simon Lee, Gene Savin, Charles Whiteman, and
Robert Wickham for their input. \ We are also grateful to J. D. Farmer for his encouragement.
\end{document}